\documentclass[final,3p,times,twocolumn,sort&compress]{elsarticle}

\usepackage{epsfig}
\usepackage{amssymb}
\usepackage{amsthm}
\usepackage{amsmath}
\usepackage{color}
\usepackage{graphicx}
\usepackage[table]{xcolor}
\usepackage{lscape}
\usepackage{caption}
\usepackage{longtable}
\usepackage{soul}

\journal{Journal of Magnetism and Magnetic Materials}

\begin{document}
\setstcolor{red}
\begin{frontmatter}



\title{Effect of an uniaxial single-ion anisotropy on the  quantum and thermal entanglement of a  mixed spin-(1/2,$S$) Heisenberg dimer}


\author{Hana  Vargov\'a\corref{cor1}\fnref{label1}}
\cortext[cor1]{Corresponding author:}
\ead{hcencar@saske.sk}
\author{Jozef Stre\v{c}ka\fnref{label2}}
\author{Nat\'alia Toma\v{s}ovi\v{c}ov\'a\fnref{label1}}

\address[label1]{Institute of Experimental Physics, Slovak Academy of Sciences, Watsonova 47, 040 01 Ko\v {s}ice, Slovakia}
\address[label2]{Department of Theoretical Physics and Astrophysics, Faculty of Science, P.~J. \v{S}af\' arik University, Park Angelinum 9, 040 01 Ko\v{s}ice, Slovakia}
\begin{abstract}
Exact  analytical diagonalization  is used to study the  bipartite   entanglement of the  antiferromagnetic  mixed spin-(1/2,$S$) Heisenberg dimer (MSHD) with the help of  negativity.  Under the assumption of  uniaxial single-ion anisotropy affecting  higher spin-$S$ ($S\!>\!1/2$)  entities only, the   ground-state degeneracy $2S$ is partially lifted and the ground state is two-fold degenerate with the total magnetization per dimer  $\pm(S\!-\!1/2)$.   It is shown that the largest quantum entanglement is reached for the antiferromagnetic ground state  of MSHD with  arbitrary half-odd-integer spins $S$, regardless of the exchange and single-ion anisotropies. Contrary to this, the degree of a quantum entanglement in MSHD with an integer spin $S$  for the easy-plane single-ion anisotropy,  exhibits an  increasing  tendency with  an obvious spin-$S$ driven crossing point. It is  shown that the increasing spin magnitude is a crucial driving mechanism for an enhancement of a threshold temperature above which the thermal entanglement vanishes.  The easy-plane single-ion anisotropy together with an enlargement of the spin-$S$ magnitude is other significant driving mechanism for an enhancement of  the  thermal entanglement in  MSHD.
\end{abstract}


\end{frontmatter}


\section{Introduction}
\label{s1}
One of the most fascinating feature of  quantum mechanical systems is an entanglement, referred to non-local correlations among the system constituents. A discovery of its importance in a quantum information processing~\cite{Niels} has initiated an intensive investigation of this unconventional phenomenon  in  various quantum systems. From the application point of view there naturally arises a question how to stabilize an entanglement at non-zero (ideally room) temperature, because an entanglement is rapidly reduced through  thermal fluctuations. 
The Heisenberg dimer represents a profound theoretical model for  analysing the bipartite entanglement, because its mathematical  simplicity and  quantum nature   merge together.    It was elucidated during last several years, that  the implementation of additional  stimuli, like  non-uniform magnetic field, seems to be a relevant mechanism for an accrument of entanglement stability~\cite{Wang09,Guo10,Xu,Guo14,Zhou15}.  Another mechanism for an enhancement of entanglement lies in an  anisotropy of a quantum system like the exchange anisotropy~\cite{Zhou03,Guo10,Guo14,Li,Zhou16,Cenci20}, Dzyaloshinskii-Moriya anisotropy~\cite{Li,Xu,Zhou16} or diversity of spin parity~\cite{Li, Huang,Hao}. Although all aforementioned mechanisms may enlarge the threshold temperature,  they unfortunately simultaneously reduce a strength  of the entanglement.  Inspired by the findings of Solano-Carrillo et al.~\cite{Solano1,Solano2}, which imply that  the uniaxial single-ion anisotropy may enhance the degree of thermal entanglement, we will   investigate the simultaneous effect of an uniaxial single-ion anisotropy and the spin magnitude on a bipartite entanglement of  MSHD.

\section{Model and Method}
\label{s2}
Let us consider  MSHD consisting of two different spins defined through the Hamiltonian 
\allowdisplaybreaks
\begin{align}
\hat{\cal H}&=J\left\{\Delta(\hat{\mu}^x\hat{S}^x\!+\!\hat{\mu}^y\hat{S}^y)\!+\!\hat{\mu}^z\hat{S}^z\right\}\!+\!D(\hat{S}^z)^2,
\label{eq1}
\end{align} 
 where $\hat{\mu}^{\alpha}$ and $\hat{S}^{\alpha}$ ($\alpha\!=\!x,y,z$) denote  spatial components of  spin-1/2 and spin-$S$ ($S\!\geq\!1$) operators, respectively. The parameter $J$ is the Heisenberg exchange interaction, $\Delta$ is the XXZ exchange  anisotropy and $D$ is the uniaxial single-ion anisotropy, which affects   higher-spin $S\!\geq\!1$ only.
 The  complete spectrum of the eigenvalues and eigenvectors of the Hamiltonian~\eqref{eq1} has been derived in our previous paper~\cite{Varga21}, so we will list here for simplicity the final formulas for eigenenergies and eigenvectors only
\begin{align}
&\varepsilon_{\pm(S+\frac{1}{2})}\!=\!\frac{S}{2}(J\!+\!2DS)\rightarrow
\vert \pm(S\!+\!1/2)\rangle\!=\!\vert 1/2,S\rangle,
\label{eq2}\\
&\varepsilon^{\mp}_{S^z_t}=-\frac{P_{S^z_t}}{4}\!\mp\!\frac{1}{4}\sqrt{R_{S^z_t}^2\!+\!Q_{S^z_t}}\label{eq3}\\
&\hskip1.cm\rightarrow\;
\vert\left({S^z_t}\right)_{\mp}\rangle=c^{\mp}_{S^z_t}\vert1/2,S^z\rangle\!\mp\! c^{\pm}_{S^z_t}\vert \!-1/2,S^z\!+\!1\rangle,\nonumber
\end{align} 
where $S^z\!=\!-S,\dots,S$ and $S^z_{t}\!=\!S^z\!+\!\mu^z\!=\!-S\!-\!1/2,\dots,S\!+\!1/2$. The coefficients $P_{S^z_t}$, $R_{S^z_t}$, $Q_{S^z_t}$, $c^{\pm}_{S^z_t}$, emergent in Eq.~\eqref{eq3} are defined as follows
\allowdisplaybreaks
\begin{align}
&P_{S^z_t}=(J\!-\!2D)\!-\!D(2S^z_t\!-\!1)(2S^z_t\!+\!1),
\label{eq4}\\
&Q_{S^z_t}=(J\Delta)^2 [4S(S\!+\!1)\!-\!(2S^z_t\!-\!1)(2S^z_t\!+\!1)],
\label{eq6}\\
&R_{S^z_t}=2(J\!-\!2D)S^z_t, c_{S^z_t}^{\mp}\!=\!\frac{1}{\sqrt{2}}\!\sqrt{\!1\!\mp\!\frac{R_{S^z_t}}{\sqrt{R_{S^z_t}^2\!+\!Q_{S^z_t}
}}\hspace*{0.3cm}}.
\label{eq7}
\end{align}
To study the bipartite entanglement we will use the concept of negativity~\cite{Peres, Horodecki,Vidal},  which is defined in terms of  negative eigenvalues $\lambda_i$ of the partially transposed density matrix $\hat{\rho}^{T_{1/2}}$, 
${\cal N}\!=\!\sum_{\lambda_i<0} |\lambda_i|$.
The zero value of the negativity  corresponds to separable (disentangled) states, whereas its  non-zero value corresponds to inseparable (entangled) ones. For MSHD, the negativity of the  maximally entangled state  equals to one-half (${\cal N}\!=\!1/2$). Detailed derivations of the negativity of  MSHD  can readers find in Ref.~\cite{Varga21}.

\section{Results and discussion}
\label{s3}
\begin{figure}[t!]
{\includegraphics[width=.47\columnwidth,trim=1cm 9.5cm 4cm 7.5cm, clip]{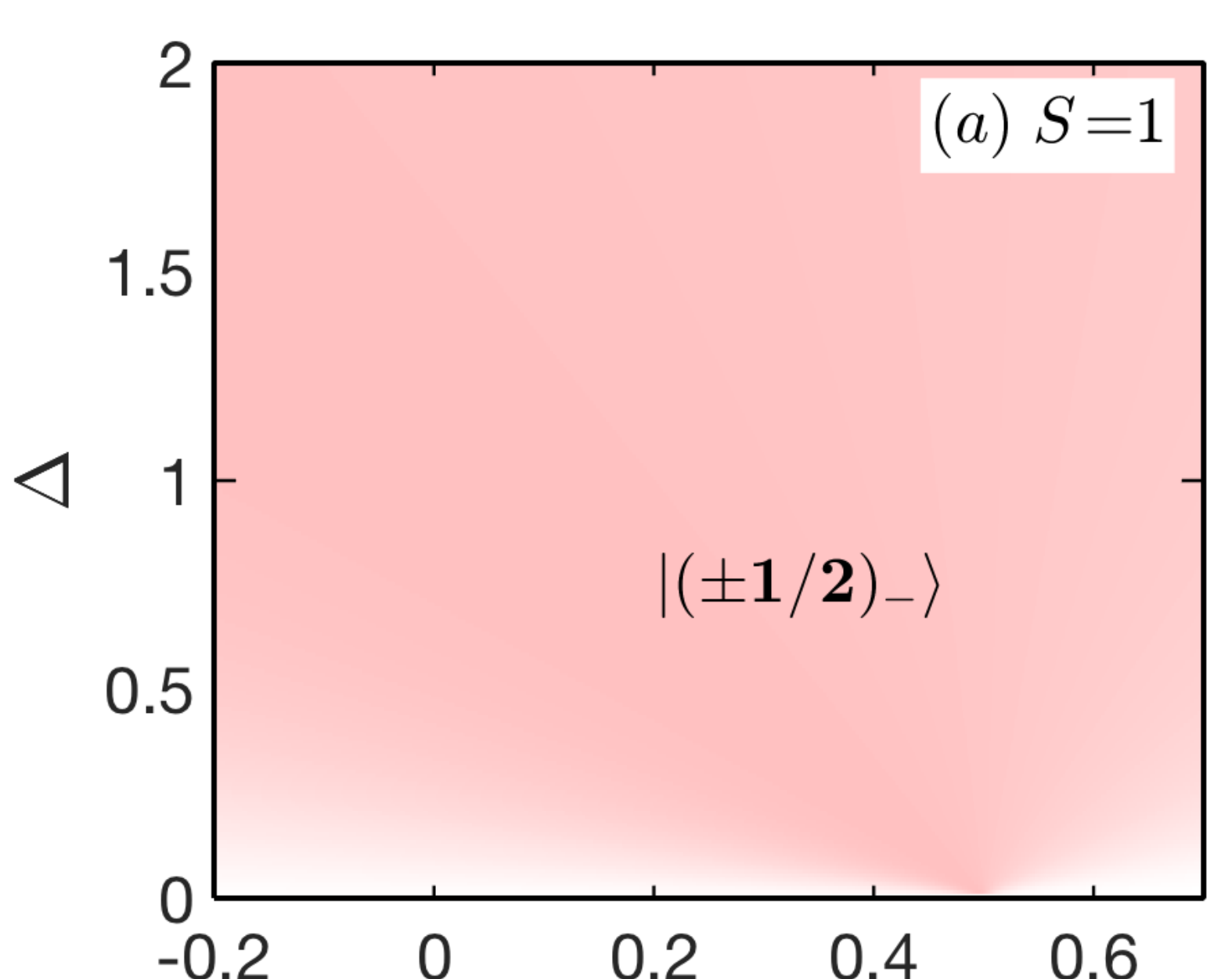}}
{\includegraphics[width=.52\columnwidth,trim=2.3cm 9.5cm 1cm 7.5cm, clip]{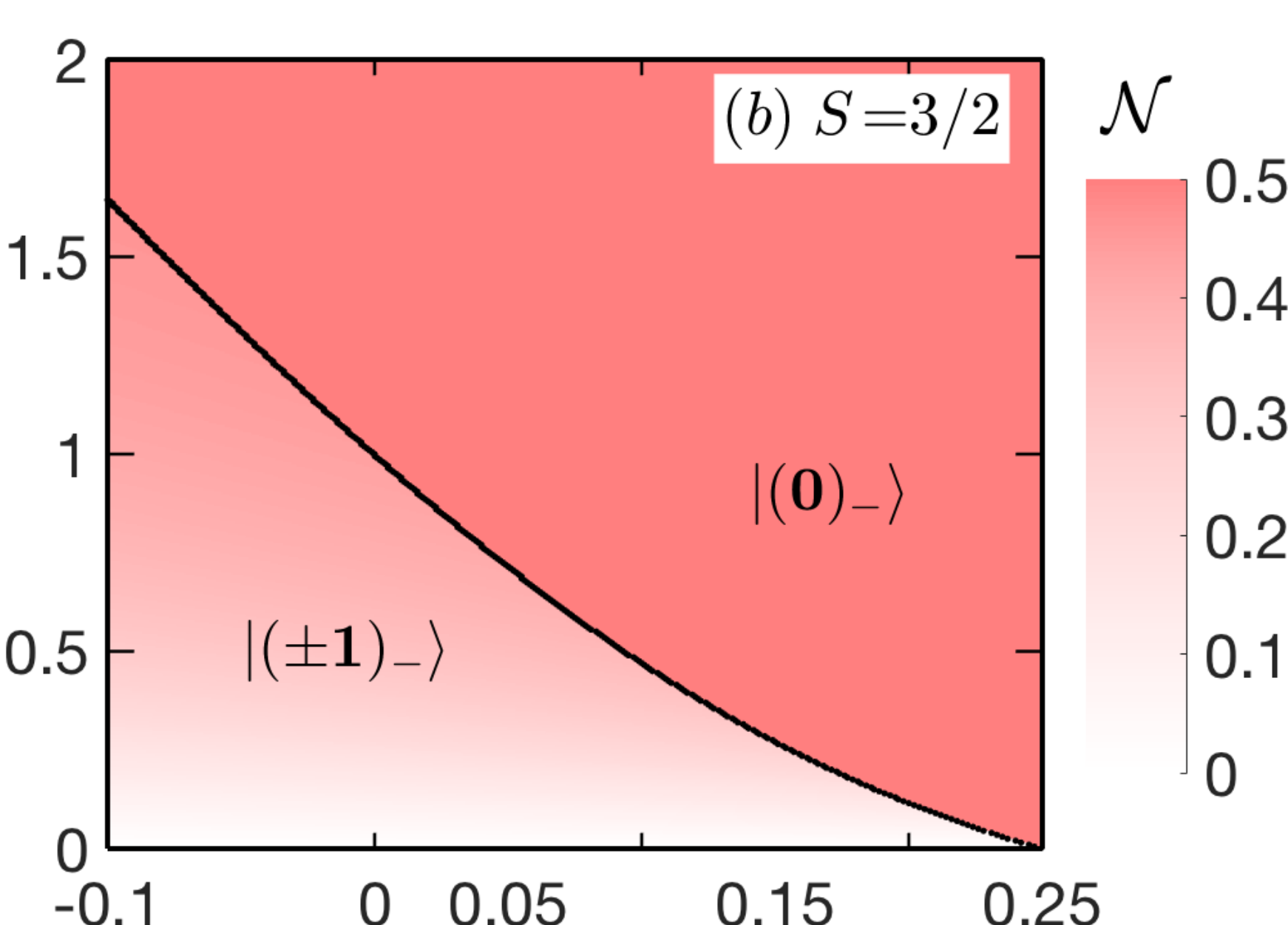}}\\
{\includegraphics[width=.47\columnwidth,trim=1cm 9.45cm 4cm 7.5cm, clip]{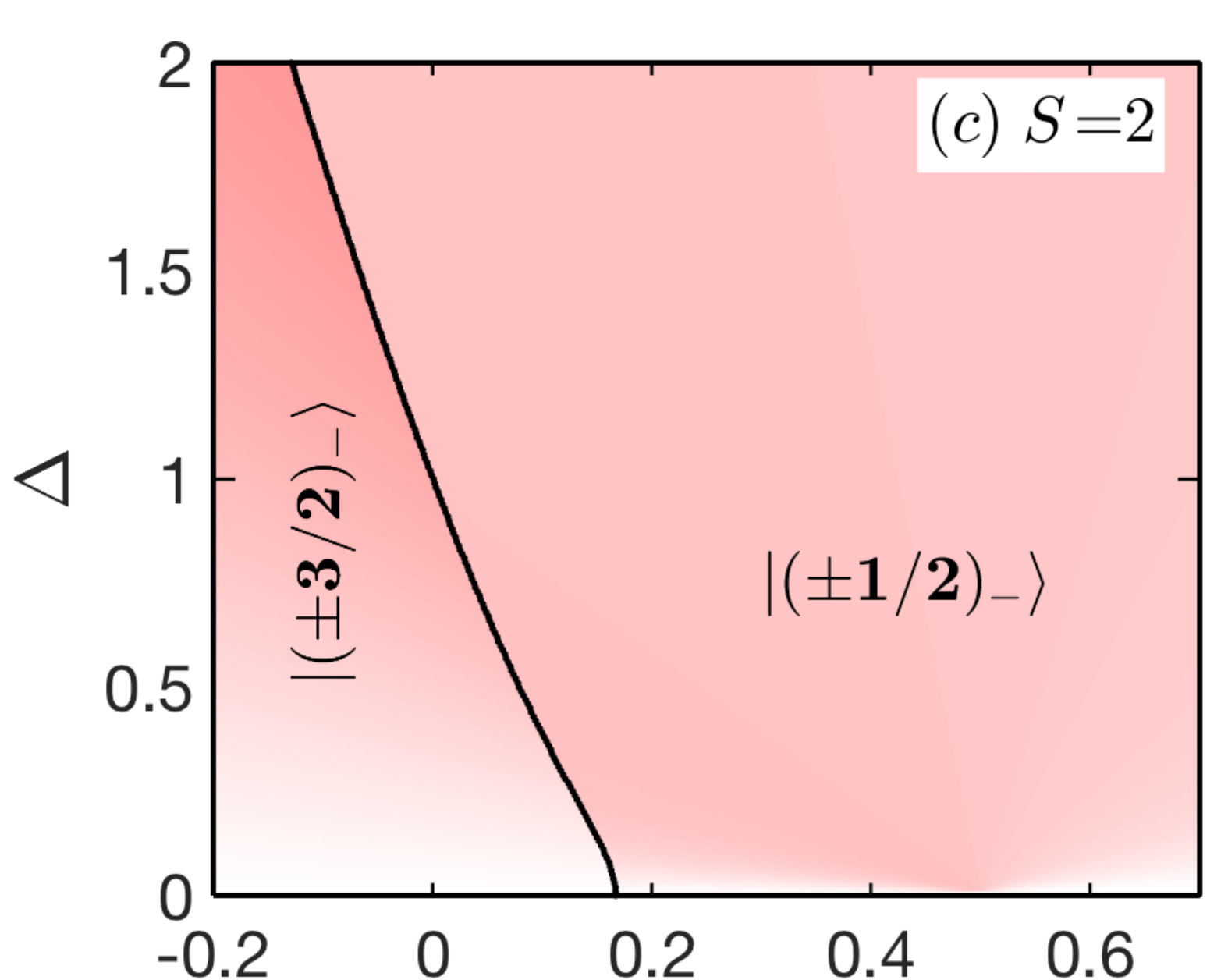}}
{\includegraphics[width=.52\columnwidth,trim=2.3cm 9.45cm 1cm 7.5cm, clip]{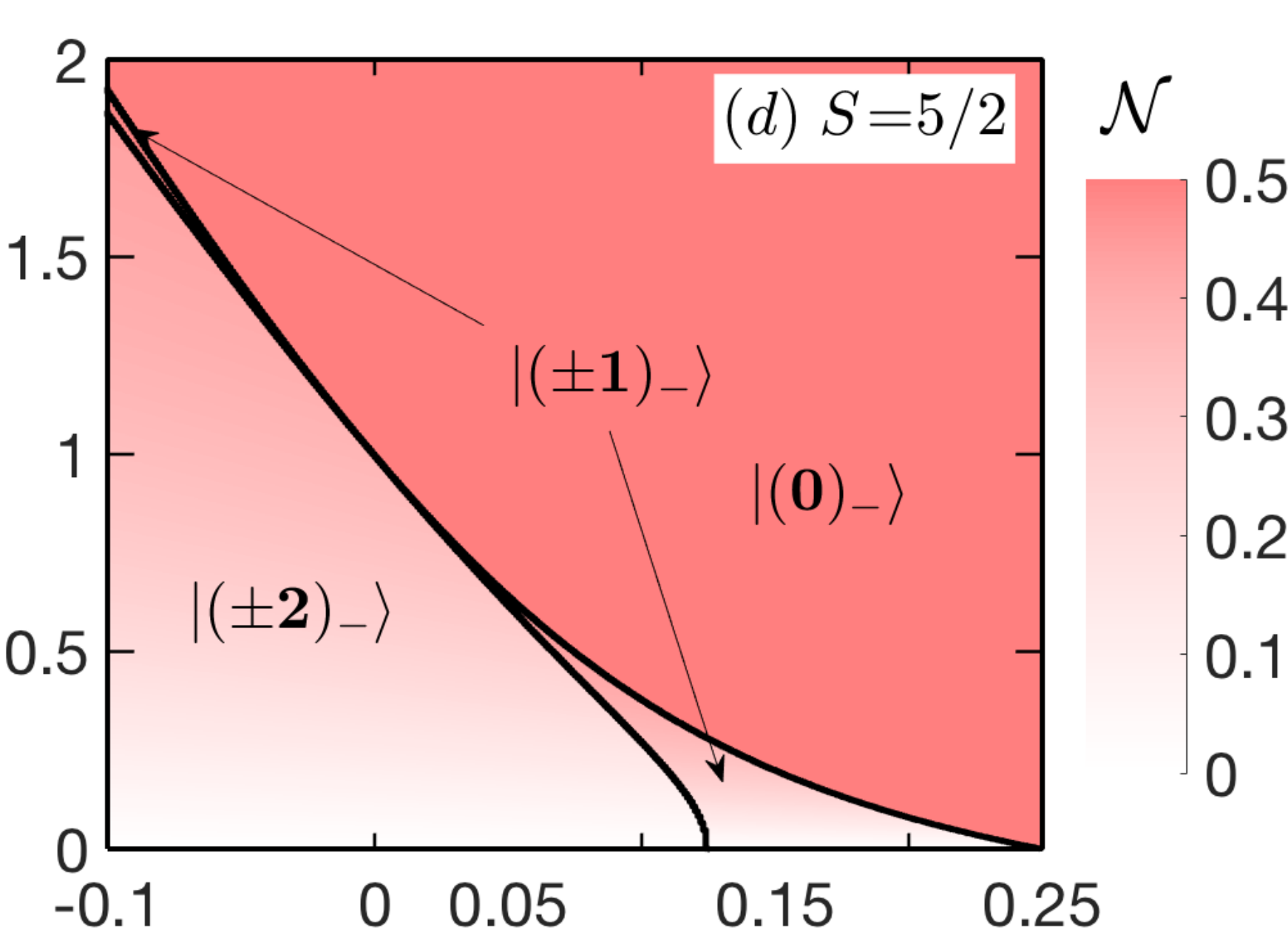}}\\
{\includegraphics[width=.47\columnwidth,trim=1cm 8cm 4cm 7.5cm, clip]{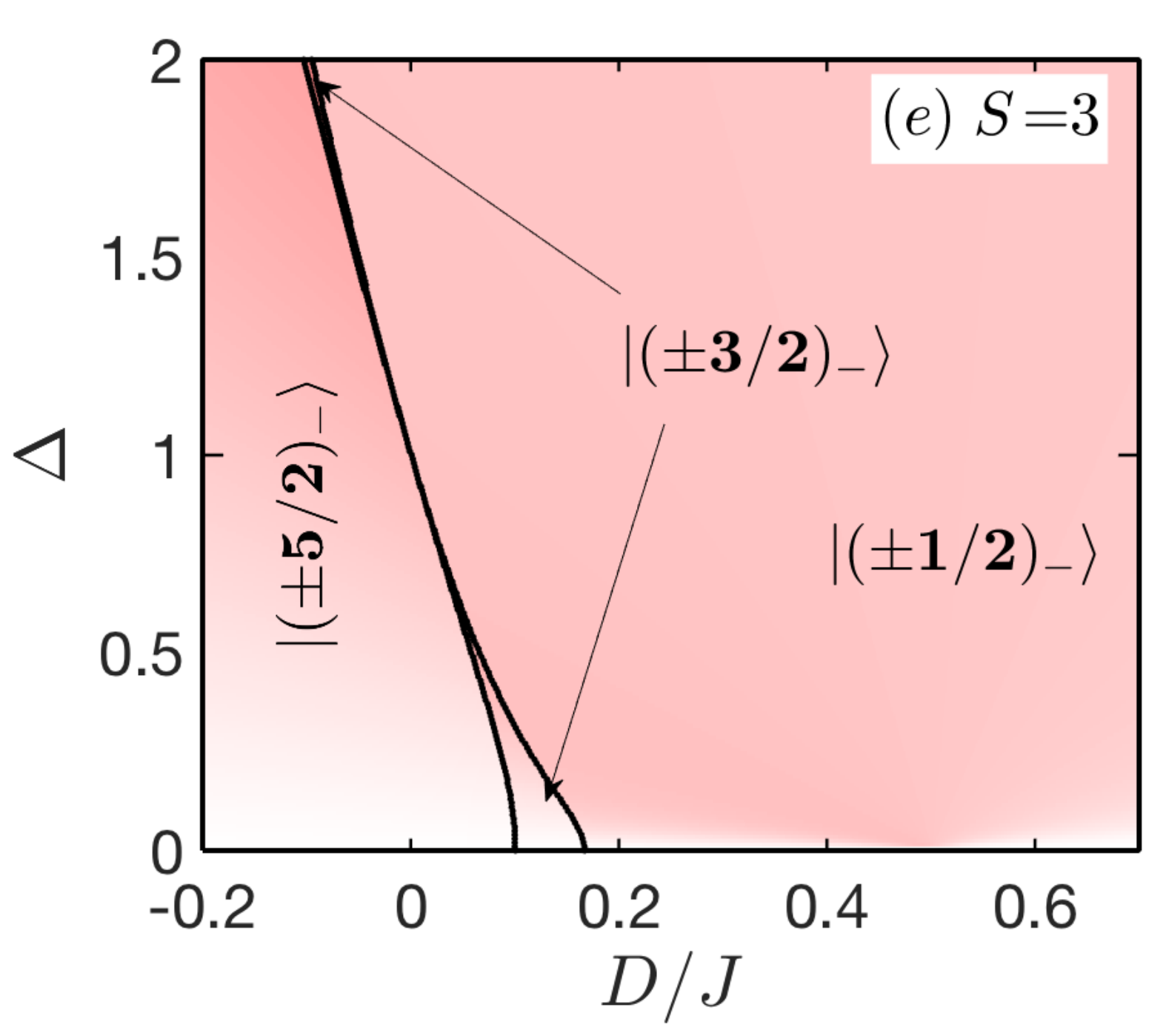}}
{\includegraphics[width=.52\columnwidth,trim=2.3cm 8cm 1cm 7.5cm, clip]{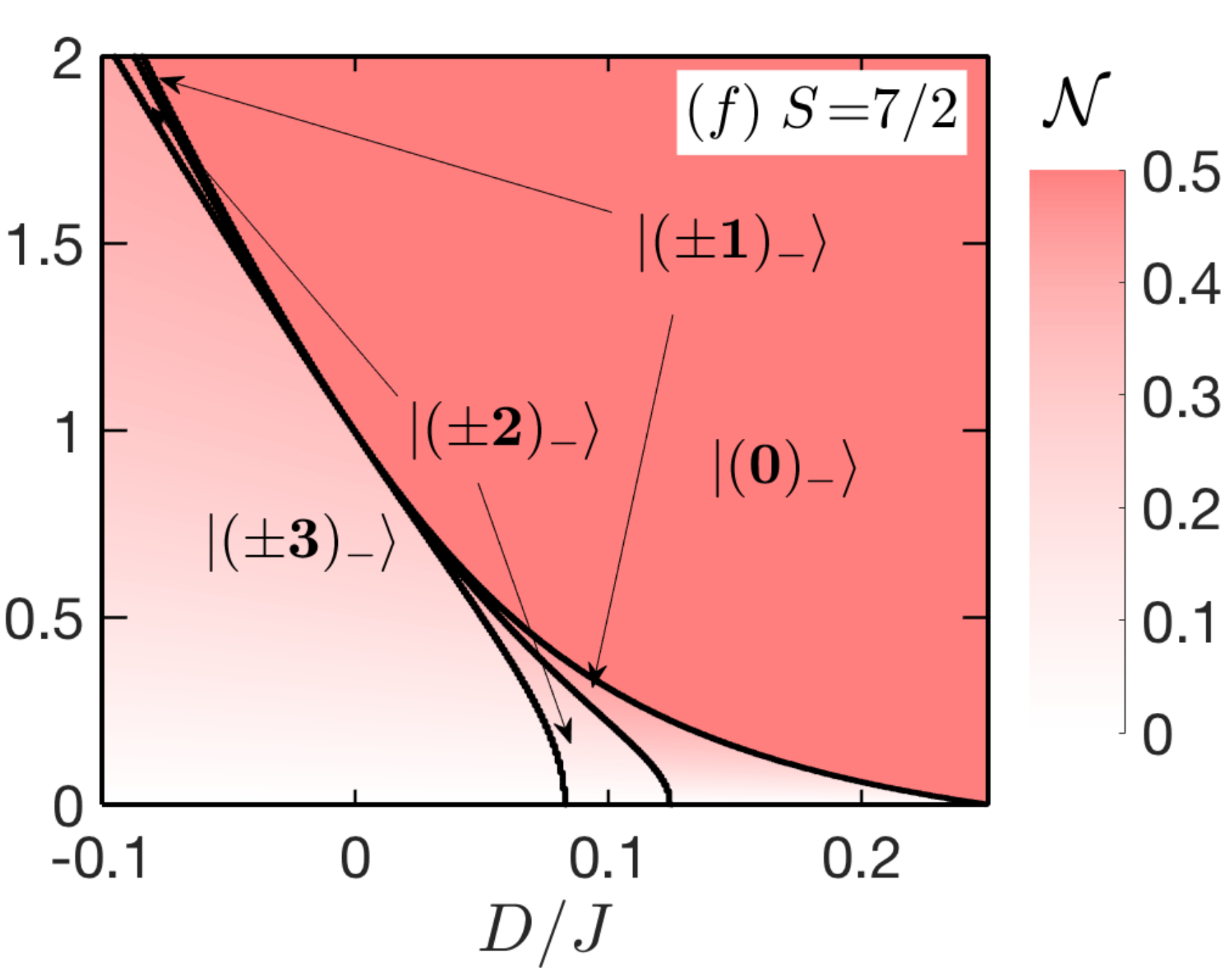}}
\vspace*{-0.6cm}
\caption{ Density plots of the negativity in the plane $D/J$ vs. $\Delta$ for integer (left panels) and half-odd-integer (right panels) spin $S$, which correspond to different ground states  $\vert (\pm S^z_t)_-\rangle$.}
\label{fig1}
\end{figure}

In the following, our  discussion will be focused on the physically most interesting antiferromagnetic case ($J\!>\!0$), for which a great variety of  ground states  is expected. Fig.~\ref{fig1} illustrates the ground-state phase diagram in the $D/J\!-\!\Delta$ plane  commonly  with the degree of entanglement exemplified through the  respective zero-temperature density plot of the negativity.  One immediately identifies that  MSHDs with both half-odd-integer constituents (right panels) remain maximally entangled if the non-degenerate antiferromagnetic ground state  $\vert (0)_-\rangle$ is favoured. A non-degenerate character of a corresponding energy level $\varepsilon_{0}^-$ determines a simple structure of the negativity, ${\cal N}\!=\!c_{0}^-c_{0}^+$, which  results in the negativity ${\cal N}\!=\!1/2$ completely independent of the strength of both  assumed anisotropies. 
It is noteworthy that  the degree of  entanglement changes discontinuously at the relevant phase boundary of the $\vert(0)_-\rangle$ ground state  and then it  continuously decreases with decreasing $\Delta$ and/or $D/J$.  For half-odd-integer spins $S$  and the  specific exchange anisotropy $\Delta$ one identifies $S\!-\!1/2$ phase boundaries, which merge together for a fully isotropic case ($\Delta\!=\!1$ and $D/J\!=\!0$)  as a consequence of energy equivalence between all $\vert (S^z_{t})_-\rangle$ ground states  ($|S^z_t|\!\leq\!S\!-\!1/2$). On the other hand, the two-fold degeneracy of each $\vert (S^z_{t})_-\rangle$ ground state  ($|S^z_t|\!\leq\!S\!-\!1/2$) for  MSHDs with integer spins $S$ (left panels) determines the  different behaviour of the negativity in the most entangled ground state  $\vert (\pm1/2)_-\rangle$. Here $\varepsilon^-_{1/2}\!=\!\varepsilon^-_{-1/2}$ and  probability amplitudes $c_{1/2}^{\pm}\!=\!c_{-1/2}^{\mp}$. Because  $S^z_t\!=\!1/2$, the probability amplitude $c_{S^z_t-1}^{\pm}\!=\!c_{1/2}^{\mp}$  and then
${\cal N}\!=\!c_{1/2}^-\left(c_{1/2}^-\!-\!\sqrt{(c_{1/2}^-)^2\!+\!(c_{1/2}^-c_{1/2}^+)^2}\right)/4$. Using  definitions \eqref{eq4}-\eqref{eq7} one obtains the relation
\begin{align}
\allowdisplaybreaks
{\cal N}\!=\!\frac{1}{4}\left( 1\!-\!\frac{1\!-\!2D/J}{\sqrt{\alpha}}\right)\left(\sqrt{ \frac{5\sqrt{\alpha}+3(1\!-\!2D/J)}{\sqrt{\alpha}-(1\!-\!2D/J)}}\!-\!1\right),
\label{eq9}
\end{align}
where $\alpha\!=\!(1\!-\!2D/J)^2\!+\!4\Delta^2S(S\!+\!1)$. 
\begin{figure}[t!]
{\includegraphics[width=1\columnwidth,trim=1cm 7.5cm 1cm 12.5cm, clip]{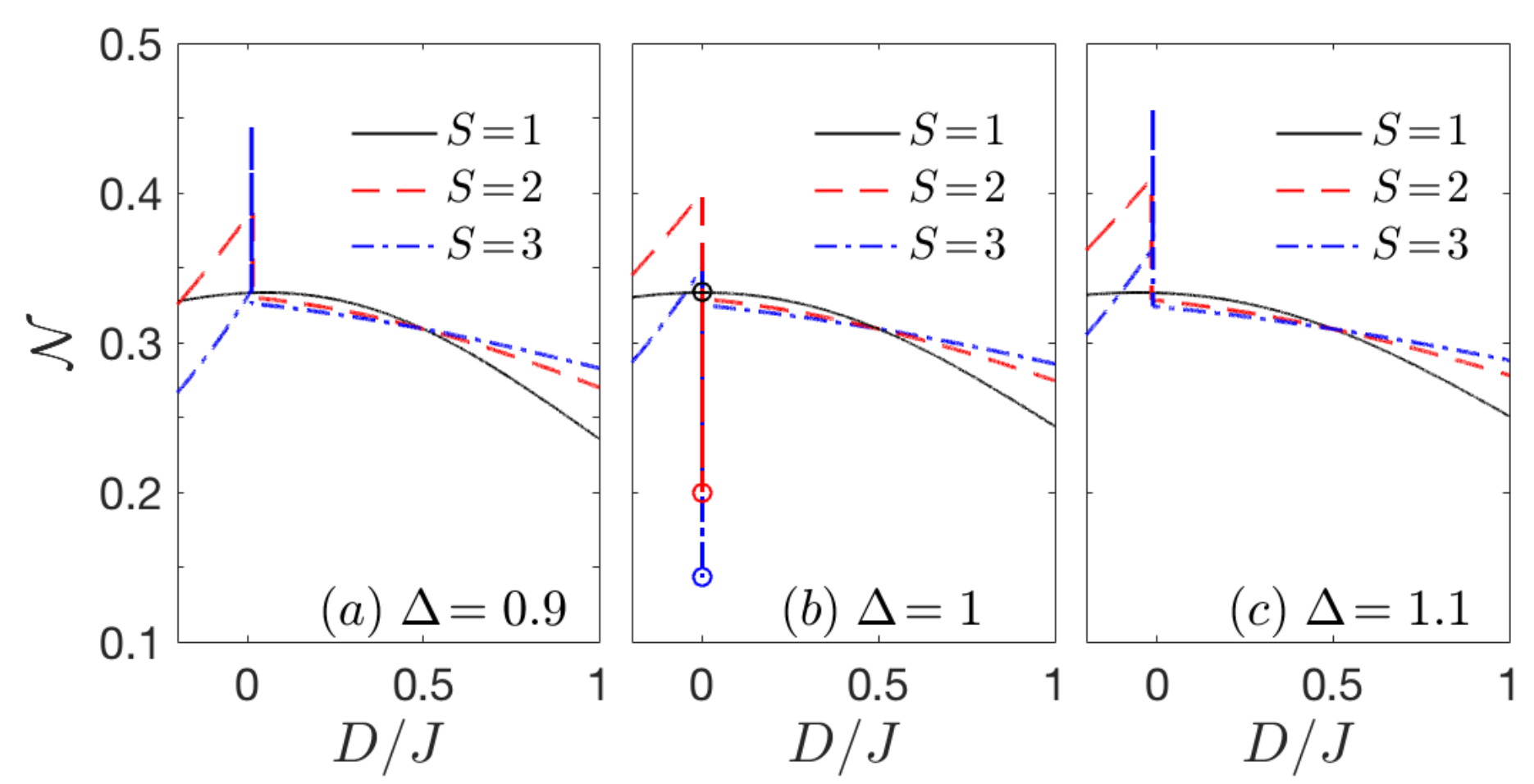}}
\vspace*{-0.2cm}
\caption{The dependence of a quantum negativity as a function of $D/J$ for different integer spin-$S$. Three exchange anisotropy limits are assumed. Open circles visualised in panel $(b)$ denote the negativity at fully isotropic case, $\Delta\!=\!1$ and $D/J\!=\!0$. }
\label{fig2}
\end{figure}
It is clear  that in the region of $\vert (\pm1/2)_-\rangle$ the degree of negativity strongly depends on the value of $\Delta$, $D/J$ and spin magnitude.  Detailed numerical analysis (Fig.~\ref{fig2}) exhibits existence of invariant negativity point at $D/J\!=\!1/2$ below which the negativity of dimer with a higher integer spin $S$ gradually decreases, whereas above $D/J\!=\!1/2$ a pronounced enhancement of entanglement is observed. 
In Fig.~\ref{fig2} one may additionally identify a singularity at a fully isotropic case ($\Delta\!=\!1$, $D/J\!=\!0$), which originates from the $2S$-fold degeneracy of   $\vert (\pm S^z_t)_-\rangle$ eigenstates. The negativity at this point rapidly falls down with  increasing spin value (open circles in Fig.~\ref{fig2}$(b)$) in accordance with the identity~(30) in Ref.~\cite{Varga21}. Because the dimer with $S\!=\!1$ can be arranged only in two-fold degenerate $\vert (\pm 1/2)_-\rangle$ ground state, the respective negativity in a fully isotropic case  is a smooth function with absence of an evident sharp minimum. Due to the same reason the negativity of  MSHD in the region of $\vert (\pm (S\!-\!1/2))_-\rangle$ behaves differently than the negativity of MSHD with  higher spins $S\!>\!1$. Focusing on the negativity at $\vert (\pm (S\!-\!1/2))_-\rangle$ phase, we confirm furthermore a previously known fact derived for $D/J\!=\!0$~\cite{Li, Huang,Hao}  that the increasing spin value $S$ ($S\!>\!1$) of  MSHD leads to the reduction of a degree of entanglement. The explanation of a different behaviour for the negativity  in $\vert (\pm 1/2)_-\rangle$  and $\vert (\pm (S\!-\!1/2))_-\rangle$  phase can be found  in an analytical expression of respective negativity functions. For $\vert (\pm (S\!-\!1/2))_-\rangle$ the negativity fulfils the relation ${\cal N}\!=\!c_{(S-1/2)}^-c_{(S-1/2)}^+\!=\!(1/2)\{(8S\Delta^2)/(8S\Delta^2\!+\!(1\!-\!2D/J)^2(2S\!-\!1)^2)\}^{1/2}$, which is evidently  inversely proportional to the spin magnitude $S$. Thus, the negativity decreases as the spin value $S$ enlarges.
%
\begin{figure*}
\centering
{\includegraphics[width=1\textwidth,trim=1.4cm 8.2cm 1.9cm 16cm, clip]{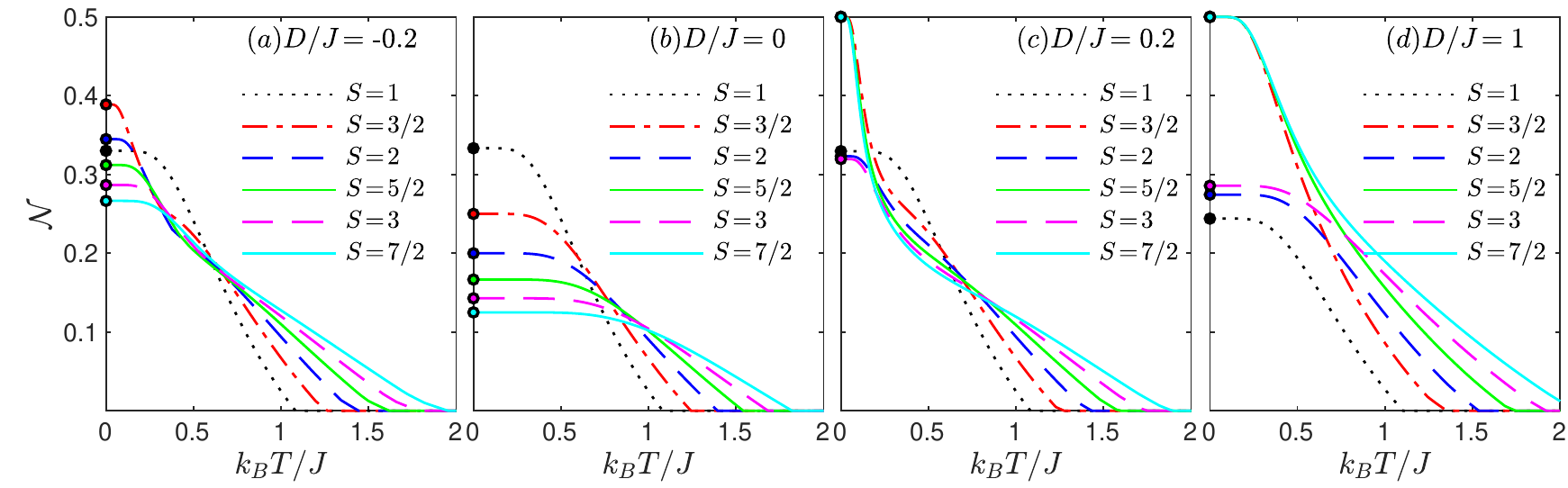}}
\vspace*{-0.6cm}
\caption{ Thermal behaviour of negativity for different spin $S$ in the mixed spin-(1/2,$S$) dimer for various  $D/J$ anisotropy.}
\label{fig3}
\end{figure*}
%

Analysing the influence of thermal fluctuations on the system entanglement we detect a few interesting observations collected in Fig.~\ref{fig3}. It should be emphasized, that  Fig.~\ref{fig3} involves  $\Delta\!=\!1$ for a simplicity,  but all  conclusions are qualitatively valid  for an arbitrary $\Delta$. (i) The increasing spin $S$ values of MSHD   always enlarges the threshold temperature above which the entanglement disappears; (ii) If the interplay between  quantum fluctuations  favours the $\vert (\pm (S\!-\!1/2))_-\rangle$ ground state (Fig.~\ref{fig3}$(a)$), 
the  negativity gradually decreases upon increasing spin size $S$ with exception of the specific case $S\!=\!1$; (iii) In the fully isotropic limit (Fig.~\ref{fig3}$(b)$), where all $\vert (S^z_t)_-\rangle$ states have an identical energy, the thermal negativity smoothly decreases upon temperature rise. Similarly as in the previous case, the higher spin $S$  is responsible for a higher thermal stability and lower degree of bipartite entanglement. Nevertheless, the highest negativity in the fully isotropic case is significantly smaller in comparison to its anisotropic counterpart due to a degeneracy of the ground state  discussed above;  (iv) The reduction effect of an increasing spin magnitude $S$ on the degree of thermal entanglement is observed equally in case of $\vert (\pm1/2)_-\rangle$ or $\vert (0)_-\rangle$ ground state  if and only if $D/J\!<\!1/2$ (Fig.~\ref{fig3}$(c)$). 
In accordance to the zero-temperature analysis, the largest negativity of  MSHDs with an arbitrary half-odd-integer spin $S$  is ${\cal N}\!=\!1/2$, whereas the negativity of  MSHD with an integer spin $S$  saturates to a smaller  value depending on both anisotropies; (iv) The stability as well as the degree of  thermal entanglement of  MSHD with $D/J\!>\!1/2$  can be enhanced by increasing  the spin size $S$. As illustrated in Fig.~\ref{fig3}$(d)$ the thermal entanglement can be additionally stabilized by  half-odd-integer spins $S$, for which  the low-temperature asymptotic limit of the  negativity achieves the maximal possible value 1/2. The increasing  
 degree of thermal entanglement driven by an increasing spin  $S$ magnitude is  also present  in  MSHDs with integer spin value $S$, however, the largest negativity does not exceed a   half of the golden ratio $(\!\sqrt{5}\!-\!1)/4$.

\begin{figure}[t!]
\centering
{\includegraphics[width=.9\columnwidth,trim=0.5cm 7.5cm 1cm 7.5cm, clip]{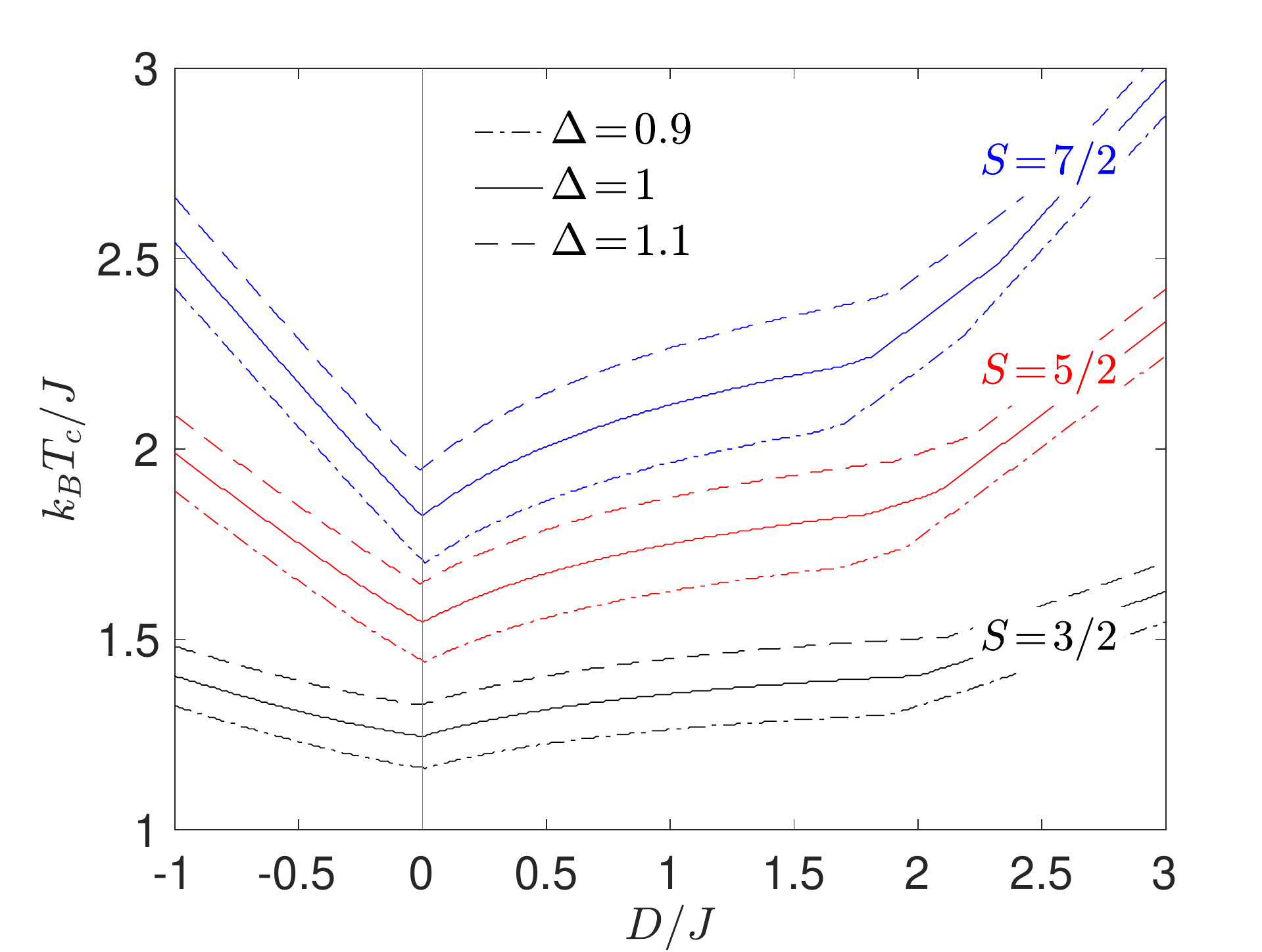}}
\vspace*{-0.2cm}
\caption{ The behaviour of the threshold temperature of the model (1) for different spin $S$ and three values of $\Delta$.}
\label{fig4}
\end{figure}

Finally, let us look at the behaviour of a threshold temperature upon the variation of  $D/J$ (Fig.~\ref{fig4}). For a better lucidity the half-odd-integer spins $S$ are assumed only.  As a result, the increasing spin value $S$ as well as the increasing  $\Delta$ enlarges the threshold temperature between the entangled and separable state.  It is evident that the single-ion anisotropy  $D/J$   has an additional enlarging effect on the threshold temperature with a significant sharp minimum at a proximity of $D/J\!=\!0$. While the  threshold temperature for  $D/J$ lying below the local minimum is a smooth continuous function, the   threshold temperature for  $D/J$ above  the local minimum involves a few pronounced flections, whose number increases as the spin magnitude $S$ increases. It is supposed that the origin of such different behaviour  relates to the different ground-state spin arrangements, where the increasing temperature affecting the starting (almost) antiferromagnetic structure can temporarily favour one of other $\vert (S^z_t)_-\rangle$ ones, until the disentangled state is achieved.
%

In conclusion, 
it was identified that the integer or half-odd-integer character of the  spin $S$  significantly determines the bipartite entanglement   both, at zero as well as non-zero temperatures.  The half-odd-integer character of both spins allows one to stabilize the maximally entangled antiferromagnetic ground state, for which the negativity is completely kept constant. On the other hand, the  negativity of  MSHDs with integer spins $S$ is not higher than ${\cal N}\!=\!(\sqrt{5}\!-\!1)/4$ and it can be tuned by magnetic anisotropies. Moreover, if $D/J\!>\!1/2$, the  negativity can be enhanced by  increasing of the spin $S$ magnitudes though it does not reach the maximal value ${\cal N}\!=\!1/2$. In addition, the  negativity in the fully isotropic MSHD ($\Delta\!=\!1$, $D/J\!=\!0$) is significantly smaller in comparison to its anisotropic counterpart as a consequence of  $2S$-fold  degeneracy. The increasing spin value $S$ rapidly  reduces its maximal value  from ${\cal N}\!=\!1/3$ achieved for the particular case with  $S\!=\!1$. 
 In  contradiction to the knowledge for $D/J\!=\!0$, the increasing spin magnitude $S$ in region of $D/J\!>\!1/2$ can be utilized to enhance not only the thermal stability but also the degree of thermal entanglement of  MSHD.  From the  application point of view,  MSHD with  half-odd-integer spin $S$ is more favourable due to its saturation to a maximally entangled state.  At the same time, the increasing spin $S$ additionally enlarges the threshold temperature between the entangled and disentangled state, which enhancement can be furthermore tuned by  uniaxial single-ion anisotropy (of the easy-axis as well as the easy-plan type).

\vspace{0.5cm}
This work was financially supported by the grant  Nos. APVV-16-0186 and VEGA 1/0105/20.

\end{document}